\documentclass[%
 reprint,
 amsmath,amssymb,
 aps,
]{revtex4-2}

\usepackage{graphicx}
\usepackage{dcolumn}
\usepackage{bm}
\usepackage{xcolor}

\begin{document}

\preprint{APS/123-QED}

\title{ Improving deep neural network performance through sampling}

\author{Lakshmi A. Ghantasala (lghantas@purdue.edu)}
 \altaffiliation[Also at ]{Ludwig Computing}
 \email{lghantas@purdue.edu}
\author{Ming-Che Li}
\author{Risi Jaiswal}
\author{Behtash Behin-Aein*}
\author{Joseph Makin}
\author{Shreyas Sen}
\author{Supriyo Datta}
\affiliation{Elmore School of Electrical and Computer Engineering, Purdue University}





\date{\today}

\begin{abstract}

Energy efficient sampling with probabilistic neurons or \emph{p-bits} has been demonstrated in the context of Boltzmann machines and it is natural to ask if these approaches can be extended to the field of generative AI where energy costs have become prohibitively large. However, this very active field is dominated by  feedforward deep neural networks (DNNs) which primarily use multi-bit deterministic neurons with no role for sampling. In this paper we first show that it is feasible to obtain superior accuracy through the use of multiple samples generated by probabilistic networks. This possibility raises the question
of which option is energetically preferable for improving
accuracy: generating more samples, or adding more bits to a single deterministic sample. We provide a simple expression that can be used to estimate these energy tradeoffs and illustrate it with results for different algorithms and architectures.


\end{abstract}

\maketitle

 \tableofcontents

 \newpage

\section{Introduction}\label{sec1}

The use of probabilistic neurons or $\it{p bits}$ comes naturally in the context of Boltzmann machines where the loss function is expressed in terms of binary stochastic variables, aka classical spins. This has given rise to the vast field of Ising computing where significant energy advantages have been demonstrated through ASICs designed to solve optimization and sampling problems (see for example Li et al. 2025 ISSCC and references therein). By contrast the field of generative AI is dominated by  feedforward deep neural networks (DNNs) which primarily use multi-bit deterministic neurons with no role for sampling.

Our preliminary results (Section 2) suggest that it may be possible to obtain comparable or even superior accuracy using multiple samples obtained from probabilistic neurons. This possibility immediately raises the question of the energy cost of generating T samples and the answer depends on various algorithmic and architectural considerations, such as the number of samples in question, the bit precision, and connective fan-out per neuron among others. 

\begin{figure}[t!]
    \centering
    \includegraphics[width=2.48in]{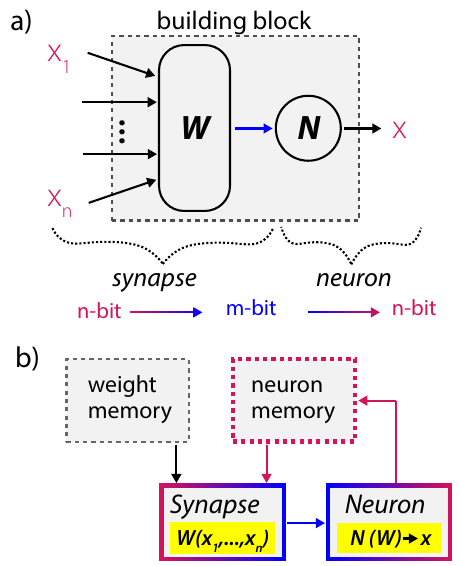} 
    \caption{(a) Building block of p-circuits, (b) System architecture used to implement the building block, based on which the energy $\varepsilon$ per elementary operation (Eq.\Ref{eq:1}) is written. }
    \label{fig:building_block}
\end{figure}

Our primary objective in this paper is to provide a simple framework that can be used to estimate these energy tradeoffs for different implementations and under different circumstances. We define a universal elementary operation, sketched in Fig. \Ref{fig:building_block}a, whose energy cost $\epsilon_{EO}$ can be written as (see Fig.\Ref{fig:building_block}b)
\vspace{-0.1cm}
\begin{align}
\label{eq:1}
\epsilon_{\text{EO}} &= 
\underbrace{n b_\text{w} \ \epsilon_{\text{wM}}}_{\text{read weights}} + 
\underbrace{(n+1) \ b_\text{a} \epsilon_{\text{aM}}}_{\text{read act.s and write out}} \nonumber \\
&\quad + \underbrace{\epsilon_{\text{S}}(n, b_\text{a}, b_\text{w})}_{\text{synapse energy}} + 
\underbrace{\epsilon_{\text{N}}}_{\text{neuron energy}}
\end{align}

\noindent where $\epsilon_\text{wM}$ and $\epsilon_\text{aM}$ represent the energy to access one bit from the weight memory and the activation memory respectively, $\epsilon_\text{S}$ is the energy needed to compute the function W over $n$ weights, $b_\text{w}$ and $b_\text{a}$ are the bits used to represent weights and activations respectively, and $\epsilon_{\text{N}}$ is the energy needed to generate a neuron output based on the output of W. The energy to write the result to activation memory is assumed equal to $\epsilon_{aM}$. 

Offhand one might assume that collecting $T$ samples would cost $T$ times as much energy, but this is not true if the samples are all generated from a single reading of the weights without reloading. In this case Eq.~\ref{eq:1} becomes
\vspace{-0.05in}
\begin{align}
\label{eq:dnn1}
\epsilon_\text{EO} =  nb_\text{w} \ \epsilon_{\text{wM}} + T \big[(n+1) \ b_a\epsilon_{\text{aM}} +  \epsilon_{\text{S}}(n, b_\text{a}, b_\text{w}) +\epsilon_{\text{N}}\big]
\end{align}
\noindent showing that the energy cost of $T$ samples can be minimal if $\epsilon_{EO}$ is dominated by the energy needed to load weights (first term).

We emphasize that we view this as a first step towards evaluating the possible role of probabilistic sampling in the machine learning space. Much depends on how many samples ($T$) are needed to achieve a significant enhancement in accuracy. The answer will clearly depend on the algorithm used to generate the samples, where the possibilities are fairly open-ended at this time. Each algorithm will then need to be  evaluated with an end-to-end evaluation framework to estimate their energy cost. Only then can we answer the question of whether for a given performance quality and energy cost it is preferable to use a deterministic $n-bit$ DNN or an $m-bit$ DNN $(m < n)$ with $T$ samples.
All we provide in this paper is (1) proof-of-concept that small values of $T$ can lead to significant improvements in performance, and (2) a framework for evaluating energy costs (Eq.\ \Ref{eq:dnn1}) with illustrative examples.


\textbf{The outline of the paper is as follows}. Section II outlines the approach used for training and running inference with p-bit infused DNNs for CIFAR10 classification. P-bit based approaches are able to \textit{sample}, which deterministic DNNs cannot do. Section III provides a common framework that may allow comparison of inference energy across a variety of DNN architectures, accounting for multi-sample probabilistic models. Section IV presents energy estimates from an FPGA implementation of a probabilistic DNN for MNIST image generation. Section V discusses some of the salient points emerging from these illustrative examples.

\section{P-bits for DNNs: P-DNNs} \label{sec3}

\begin{figure*}
\centering  \includegraphics{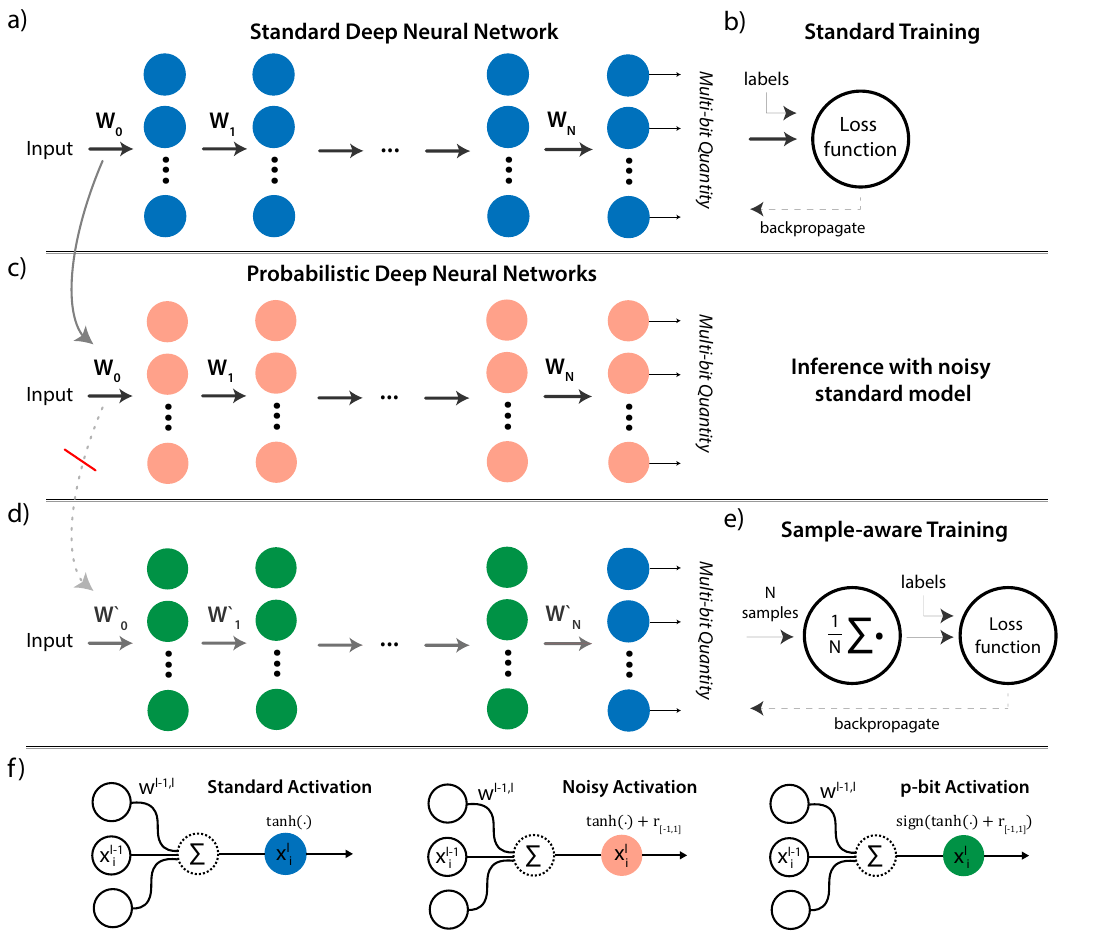} 
    \caption{(a) Standard DNNs are implemented with activation functions like tanh or ReLU, and (b) are generally trained following backpropagation. PDNNs can come in different forms, broadly referring to DNNs with stochasticity infused into each forward pass. (c) shows a model that integrates stochasticity via noisy activations, while (d) shows a model that replaces the activations with pbits. (e) shows the sample aware training scheme that dramatically improves performance for pDNNs. (f) specifies the activation functions.}
    \label{fig:overview}
\end{figure*}

Let us first discuss \textit{two possible ways} to incorporate p-bits into deep neural networks (DNNs) to construct probabilistic or p-DNNs.

 

 \subsection{Sample-aware training of DNNs}
 A standard approach to realize low-bit DNNs is through $\textit{quantization-aware}$ training, which generally is applied to weights, though the idea can extend to activations as well \cite{hubara_quantized_nodate}.
 We follow a similar approach, but instead train with multiple samples, following a \textit{sample-aware training} scheme as shown in Fig.\ \ref{fig:overview}c.
 This approach involves using the loss value computed from the average of $s$ samples to perform the training by backpropagation \cite{raiko_techniques_2015}. The idea then is to \textit{replace} the activations of a traditional DNN with stochastic activations, and then train the model to take into account the stochasticity. 

 Through sample-aware training, the multiply-accumulate operation, traditionally a computation bottleneck in DNNs \cite{capra_hardware_2020}, is simplified to being between a multi-bit digital weight and a single-bit activation. This dramatically reduces the compute energy, where compute is the sum of synapse and neuron energies, though one has to be wary that drawing multiple samples will again increase that compute energy. We will show that 2 samples is enough to improve accuracy. 
 
 We apply this sample-aware training scheme to two p-DNNs, a custom VGG5 model (see Fig.\ref{fig:model-architectures}b) for image classification trained on the CIFAR10 dataset, and a custom VAE model (see Fig.\ref{fig:model-architectures}a) for image generation trained on the CelebA dataset. 

\textbf{Image classification:} We show that one sample of a 1-bit activation p-DNN is enough to match deterministic accuracy, with 2 samples outperforming. Since this amounts to adding a minor compute energy (and \emph{not} memory energy), the overall energy cost is quite small for the accuracy bump, as seen in Fig.\ \ref{fig:energy_dnn}b. Additional samples from the p-DNN model lead to higher accuracy, roughly matching 3-bit deterministic accuracy at 10 samples as seen in Fig.~\ref{fig:acc_plots}a. The colored lines in Fig.\ \ref{fig:acc_plots}a use multi-sample training and inference by averaging \textit{at the end} of the model. The dotted black line shows averaging test-time samples \textit{at each layer} in the model, with the same weights as those in the deterministic baseline. Interestingly, while the algorithm improves accuracy even at 1-sample inference of a model trained at 45 samples for the classification task, it only improves inference at the same number of samples for the generation task, which we will discuss next.

\begin{figure}
    \centering
    \includegraphics[width=1\linewidth]{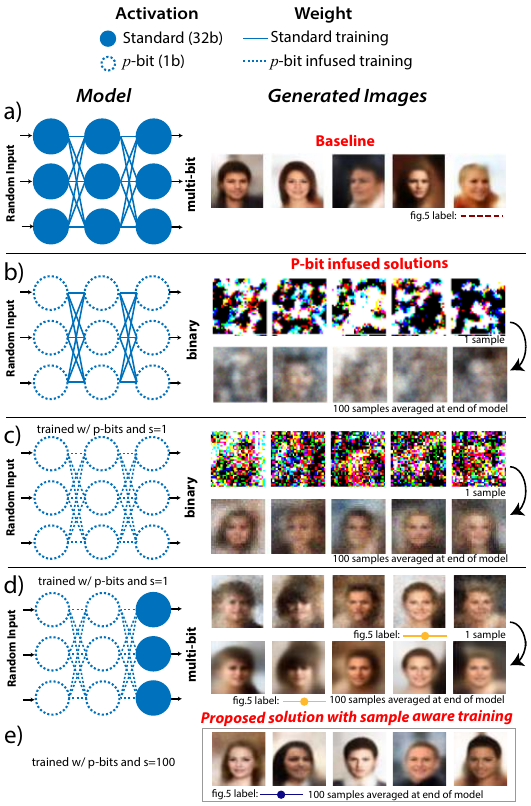} 
    \caption{Images generated via a traditional variational autoencoder compared with various probabilistic alternatives of the same architecture. (a) standard DNN inference. (b) model with pbits replacing all activations. (c) model that is retrained with pbit activations. (d) model that is retrained with pbit activations and an analog last layer. (e) proposed sample aware training scheme employed on (d). }
    \label{fig:image_gen}
\end{figure}

\textbf{Image Generation:} As the baseline, we use a 12-layer Variational Autoencoder DNN using 32-bit weights and sigmoid activations trained to generate celebrity images from a random input (Fig.\ \Ref{fig:image_gen}a).
Replacing the 32-bit activations with p-bits we get very noisy images, but averaging over 100 samples creates a hint of recognizable facial images (Fig.\ \Ref{fig:image_gen}b).
Retraining the weights using the actual nonlinear element instead of the continuous version, the quality of images improves significantly (Fig.\ \Ref{fig:image_gen}c). Replacing the p-bit activations in the final layer with continuous sigmoid activations marks an improvement, even at 1 sample (Fig.\ \Ref{fig:image_gen}d, top row).
High-quality images comparable to the original baseline are obtained (see final row of Fig.\ \ref{fig:image_gen}d) with sample-aware training. Fig.~\ref{fig:image_gen} compares the image quality for different schemes quantitatively using the Fr\'echet-Inception Distance (FID). A key point to note is that better performance is obtained when the number of samples used in testing \textit{matches} that used in re-training. Furthermore, with sample-aware training, images generated with p-bits can reach very close to 32b FID scores.

Surprisingly, even simple averages at each p-bit output using weights from a 32b deterministic model seem to produce recognizable face images at just 100 samples, as opposed to the expected $O(2^{64})$. It would seem that though standard error applies for \textit{each operation}, it doesn't necessarily apply to the \textit{end application} (e.g.\ image generation), which has some robustness built into it through the training process. This extends to classification problems as well, as we'll discuss next. 

Implementation details for both experiments are included in appendix \ref{App:simulation_details}.


\subsection{Sampling noisy DNNs}\label{sec:sampling_noisy_dnns}


While sample-aware training requires re-training a model with stochastic activations, there may be ways to take advantage of sampling on DNNs even without retraining, through added noise.
By adding some noise to a deterministically trained model, it is possible to generate `noisy samples' that when averaged, produce lower loss results than a standard model.  

\begin{figure*}
    \centering
    \includegraphics[width=7in]{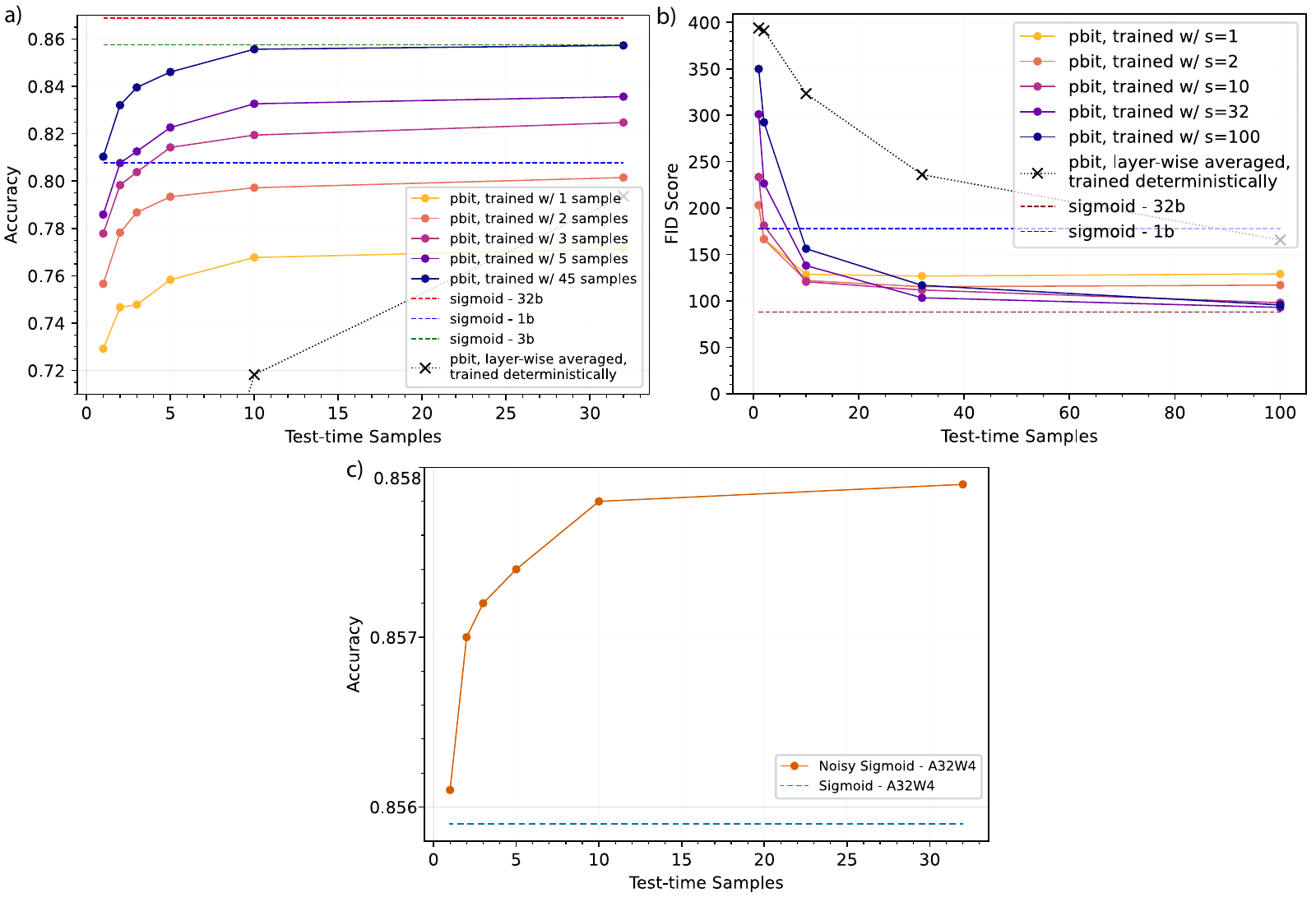} 
    \caption{(a) Comparison of accuracy for CIFAR10 classification.  (b) Comparison of FID metrics for variational autoencoder trained on the celeba dataset. (c) Accuracy on CIFAR10 of a deterministic baseline compared with model with noisy activations.}
    \label{fig:acc_plots}
\end{figure*}

We apply this approach to the activations of a 32b model trained for CIFAR10 classification.
The model is run at 4b in a standard post-training quantization fashion \cite{jacob_quantization_2018} with group size 64 , as well as quantized after adding some noise to the original model. With just 2 samples, the sampled model produces better results, shown in Fig.\ \ref{fig:acc_plots}c. The improvement shown here is modest, but what is surprising is that the improvement is monotonic. One might have expected that adding noise to a deterministically trained model would randomly improve or degrade performance with samples, but that does not seem to be the case.


We've now discussed two potential approaches to taking advantage of sampling via p-DNNs:\ a model trained with p-bit activations, and a model with noise added to standard activations. In problems of this type, our aim is not to recreate the activations $A$ precisely but rather to minimize the loss function $\mathcal{L}$ characterizing the mapping of probability distributions. This is facilitated by the fact that, at the optimum, the partial derivatives $\frac{\partial \mathcal{L}}{\partial A}$ are on average zero, so the loss is insensitive to small perturbations in the activations. We can get better results by taking into account more samples to recover any loss of $\mathcal{L}$ due to the noise we add to $A$. In the next section, we will discuss the energy cost of each implementation and how they (and potentially any other method) fit into the energy framework outlined by Eq.\ \ref{eq:1}.  

 


\subsection{Contrast with stochastic bitstreaming}
It should be noted that using p-bits in place of continuous neurons as described above is \textit{very} different from the well-known use of stochastic bitstreams to represent continuous signals. With stochastic bitstreams the aim is to generate streams of single bits whose mean value matches the original 32-bit value. Since the mean of a non-linear function is not equal to the function of the mean, sample averaging should be performed individually for \textit{each linear operation at each layer}. Incidentally, we note that the stochastic bitstreaming approach does achieve results that exceed statistical expectations derived from its standard error scaling, though not enough to match the presented mechanisms. 



\section{Energy analysis of sampling systems} \label{sec2}

The solution could be reached faster at the same power level if the energy per elementary operation $\epsilon_\text{EO}$ could be lowered, for example through the use of in-memory computing and/or analog addition followed by a sampler that uses the analog sum to generate a binary output. These are part of the hardware design while the quantities $N$, $n$ and $T$ are set by algorithmic considerations. Let us elaborate a little on the different energy components of the elementary operation described by Eq.\ref{eq:1}: \\

\subsection{The core components}

\indent \textbf{\textit{Memory access energy}}, $\epsilon_{wM}, \epsilon_{aM}$: 
Most large scale applications in machine learning are memory bottle-necked in latency and energy, making this component especially valuable in an end-to-end analysis of probabilistic solutions. In \cite{li_122_2025} both weights and p-bit values were stored in local registers, making their read energies identical and relatively low in the tens of fJ range. With large problems it may not be possible to store all weights in registers, requiring SRAM or even DRAM with much larger read energies. Separating weight and activation memory is useful in that p-bits as activations are binary, and may be stored in closer-to-compute memory than weights, which may be orders of magnitude larger overall. In that case, in-memory computing options can be employed to minimize read energies, which can otherwise dominate the other components (Fig.\ref{fig:energy_dnn}). In Eq.\ref{eq:1}, memory access energy involves reading $n$ weights, reading $n$ activations, and writing the output activation of the building block back to memory. 

\textbf{\textit{Synapse energy}}, $\epsilon_{S}(n, b_w, b_a)$: 
The function W is usually linear: $w_1x_1 + w_2x_2+ \cdot \cdot + w_nx_n$ for $b_w$-bit $w_i$ quantities. Usually a multiply-accumulate unit (MAC) is used to evaluate W, but for p-bit activations, since ${x_1, \cdot \cdot x_n}$ become binary variables, the function $W$ only requires an accumulate (AC) function that selectively sums a subset of the weights ${w_1, \cdot \cdot w_n}$. Based on \textit{n}, there may be architectural considerations to take into account. For instance, the problems in \cite{li_122_2025} use $n=6$, which is relatively small so that the function W can be read off a look-up table (LUT) with $2^n = 64$ entries. For p-bit building blocks where $n$ is too  large for the compute to be implemented with a LUT, $\epsilon_S \sim n~\epsilon_{add}(b_w)$, where $\epsilon_{add}(b_w)$ is the energy for adding two $b_w$ bit numbers via an adder. For deterministic building blocks, $\epsilon_S \sim n~ \epsilon_{mul}(b_a, b_w) + \epsilon_{add}(b_a)$, where $\epsilon_{mul}$ is the energy for multiplying a $b_a$ bit neuron with a $b_w$ bit weight. These products then need to be accumulated with the add.  

\textbf{\textit{Neuron energy}}, $\epsilon_{N}$:
The neuron energy is defined as the energy needed to apply some non-linear function to an input. For p-bit based systems, the neuron may be implemented following the function
$b = \text{sign} \big(\tanh(W)-\text{rand}_{\{-1,+1\} }\big)$
\noindent where rand$_{\{-1,+1\}} $is a random number distributed uniformly between $-1$ and $+1$. The p-bit based neuron energy then has three sub-components: (1) compute the activation ($\epsilon_{\text{act}}$), (2) generate $\text{rand}_{\{-1,+1\}}$ ($\epsilon_{\text{RNG}}$), and (3) compare the two to generate a binary output ($\epsilon_\text{compare}$). Note that unlike the memory read and compute energies, the sampling energy for p-bit based systems is independent of $n$ and becomes relatively unimportant for large values of $n$, i.e the neuron fan-in. In deterministic systems, there would be no RNG; for instance, in machine learning models, the neuron generally takes the form of some activation like ReLU or swiglu. For more complicated models, the neuron may take the form of softmax, or normalization, which we will discuss later. 

We define these components not to say that all probabilistic solutions should map to a single rigid formula, but rather to establish a framework with which to build custom energy functions for unique situations. 

Fig.\ref{fig:energy_qmc} reveals that for low-K Boltzmann machine architectures,  compute energy dominates the overall system. This motivates replacing the compute unit in these systems, which naturally requires sampling, with real p-bit hardware like s-MTJs or zener diodes to achieve meaningful end-to-end energy efficiency improvements. We will show in the coming sections that DNNs are not compute dominated, but rather memory dominated. We will argue that in such systems, minor increases in compute are almost negligible in the end-to-end pipeline in terms of energy, but can have a profound impact on the accuracy metrics of the task. 

\subsection{65nm ASIC for Quantum Monte-carlo}
\begin{figure}
    \centering
    \includegraphics[width=1\linewidth]{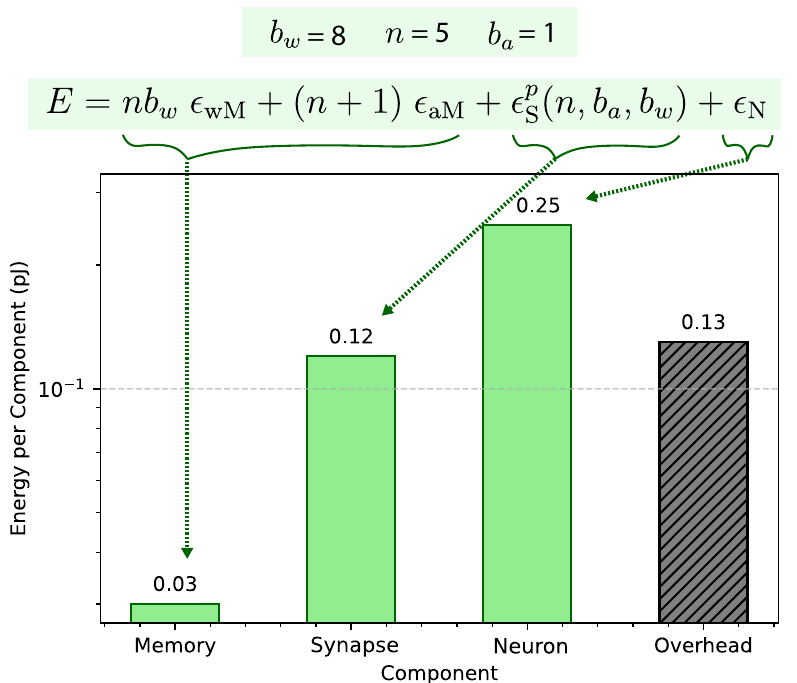} 
    \caption{Energy cost of component operations of building block for QMC from 65 nm ASIC implementation reported in Ref.\cite{li_122_2025}.}
    \label{fig:energy_qmc}
\end{figure}

Let us now discuss how the energy equation applies to a standard Boltzmann machine implemented with 65 nm CMOS technology  (Ref.\cite{li_122_2025}) which was used to solve a diverse set of problems having very different values of $N$, problem size, and $T$, sample requirement, but in each case the energy per elementary operation was about the same: $\epsilon_p \sim 0.5 pJ$. The energy split between the components of Eq.\ref{eq:1} is depicted in Fig.\ref{fig:energy_qmc} for a specific configuration of hyper-parameters pertaining to the QMC problem solved in \cite{li_122_2025}.  

A fourth energy component acting as a design-specific catch-all term is added to Fig.\ref{fig:energy_qmc} to include any control logic, I/O, clock distribution, routing, or other overheads. For the ASIC in \cite{li_122_2025}, this design-specific overhead amounted to 25\% of the overall energy consumption, but we have not included this component in Eq.\ref{eq:1}, since it is often not fundamental to the underlying algorithm or architecture. 

Fig.\ref{fig:energy_qmc} shows that the quantum-monte carlo problem is dominated by compute energy, where we can define compute as the sum of synapse and neuron energies. In such systems, it is clear that introducing additional compute would require careful consideration. In the next example, we will show that additional compute is much less costly for DNNs, and can be taken advantage of for notable accuracy gains. 

\subsection{DNN's with sampling, p-DNN's}


\begin{figure*}
\centering
\includegraphics[width=1\linewidth]{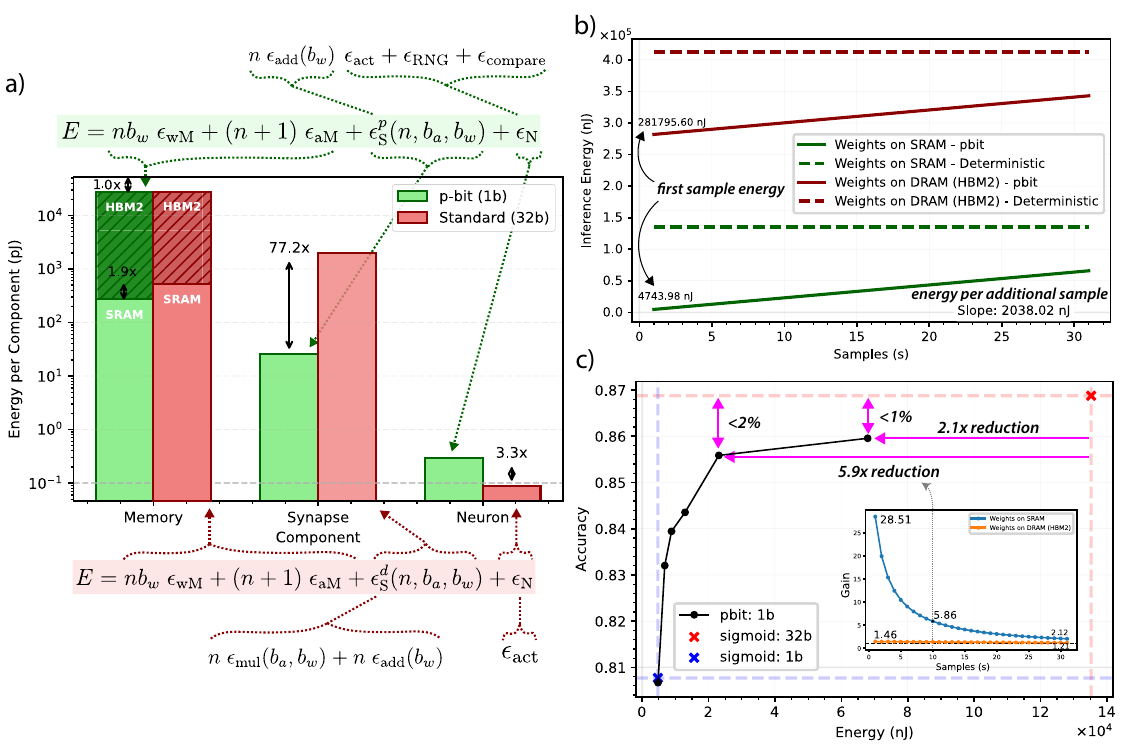}
\caption{(a) Energy of component operations of building block for DNN p-bit and deterministic building blocks. $n$ is the average number of inputs to a nonlinearity across all layers of the image generation DNN, scaled to match dimensions of ResNet50. The SRAM components use $\epsilon_{aM} = \epsilon_{wM} = 0.0384$ pJ/bit based on 1024 word, 144 bits/word access to 65nm TSMC process. The HBM2-DRAM components use $\epsilon_{aM} = 3.97$ pJ/bit \cite{oconnor_fine-grained_2017}, while retaining the activations on SRAM. (b) shows energy per sample for the same setup as described for (a). (c) combines energy per sample with accuracy per sample to show accuracy vs energy for the same setup as described for (a).}
\label{fig:energy_dnn}
\end{figure*}






\noindent The energy $E$ for an N-neuron DNN is $N$ times the energy per elementary operation $(\epsilon_{EO})$ in Eq.~\ref{eq:dnn1}:
\begin{equation}
\label{eq:dnn}
E = N \big( nb_w \ \epsilon_{\text{wM}} + T(n+1) \ b_a\epsilon_{\text{aM}} +  T\epsilon_{\text{S}}(n, b_a, b_w) + T \epsilon_{\text{N}} \big)
\end{equation}
where a p-bit implementation could average over T samples to improve prediction quality. We bring T inside the elementary operation as multiple samples can be generated without reloading the weights. This is why drawing multiple samples is significantly cheaper relative to drawing a single sample, where the weight loading energy cannot be amortized. 

We can consider two cases for Eq.\ref{eq:dnn}, a p-bit case and a standard case, where the p-bit case mimics the sample-aware training models (i.e activations are replaced with p-bits). For the p-bit case, $\epsilon_N \rightarrow \epsilon_\text{act}+\epsilon_\text{RNG} + \epsilon_\text{compare}$ and $\epsilon_S \rightarrow n \ \epsilon_\text{add}$, while for the deterministic case, $\epsilon_N \rightarrow \epsilon_\text{act}$ and $\epsilon_S \rightarrow n \ \epsilon_\text{mul} + n \ \epsilon_\text{add}$. The energy gain can then be defined as $\text{gain} = \frac{E_{d}}{E_p}$ where $E_d$ is the energy draw for the deterministic DNN and $E_p$ for the p-DNN. The $n$ used is the average fan-in weighted by the number of parameters per layer in the VGG5 model in Fig.\ref{fig:model-architectures}b. The standard DNN building block multiplies $n$ weights (each $b_w$ bits) with $n$ activations (each $b_a$ bits) while the p-bit implementation only sums up a selected subset of $n$ weights (each $b_w$ bits), corresponding to activations with $b_a$=1. 

By plugging in the values for these quantities, recorded in Tab.\ref{tab:energy_eq_variables} and visualized in Fig.~\ref{fig:processing_element}, we can estimate the overall energy draw of a custom hardware built to run inference using these models, which is shown in Fig.\ref{fig:energy_dnn}. These quantities are simulated using the same platform and PDK used to design the ASIC in \cite{li_122_2025}. Note that this analysis compares p-DNNs with DNNs at the same weight bit width.

The following are some critical insights we can gather from this experiment.
\begin{itemize}
    \item In contrast to the QMC chip results in Fig.\ref{fig:energy_qmc}, the memory energy dominates the overall ML pipeline, even for small sized models ($\sim$2M parameters), as seen in \ref{fig:energy_dnn}a. For applications requiring DRAM, which are most production scale LLMs today (100B+ parameter models), adding some compute for additional performance would be very desirable, as it would have little impact on the end-to-end energy cost of inference. 
    \item As shown in Fig.\ref{fig:energy_dnn}a, p-DNNs using p-bit activations show dramatic improvement in synapse energy owing the reduction of activations to 1 bit, the MAC to AC benefit.  
    \item The first sample from a probabilistic DNN will be more costly than subsequent samples (see Fig. \ref{fig:energy_dnn}b.). The energy cost of loading the weights from memory are amortized across multiple samples, but are felt entirely for 1-sample inference. Intermediate activation tensors will be stored back in activation memory in a $T \times b_a$ memory chunk, and each layer will overwrite the same memory with new activation samples. 
    \item Fig.\ref{fig:energy_dnn}c shows that tolerance for a 1\% accuracy drop-off can come with 2x+ reduction in energy used going from a 32b DNN to p-DNN. We use accuracy per number of samples from the VGG5 experiment in Fig.\ref{fig:acc_plots}.a and combine it the energy per sample values in the inset of Fig.\ref{fig:energy_dnn}c to show an energy vs accuracy curve comparing probabilistic and deterministic DNNs. From a purely compute perspective, though quantized models (like the 3b model matching performance of 10 p-bit samples) are compitetive in energy iso-accuracy to stochastic models, it is important to remember that p-DNNs are \textit{run-time adjustable}. More or less samples can be drawn at run-time to fit the application at hand, dynamically adjusting energy and accuracy to suit the user's needs. 
\end{itemize}

\subsection{Expanding on the framework}
The real power of this framework is how easily it can be tweaked and expanded on as needed. For example, this framework can be tweaked to find the energy per inference for an entire model, even for models that may not traditionally be thought of as breaking down neatly into \textit{synapse} + \textit{neuron} elementary operations. For instance, Large Language Models (LLMs) like Llama have various nonlinearities in addition to more traditional neurons like the SiLU, including but not limited to RMSNorm, softmax, and ROPE. The last term ($\epsilon_N$) of Eq.\ref{eq:1} can be split into multiple neuron types as needed to cover these various nonlinearities. The energy for the entire model can be defined by scaling weight reads by the number of weights across the model, scaling synapse energy by the number of macs across the model, and scaling each neuron type's energy by the number of those neurons in the overall model. 

From Eq.\ref{eq:1} then, one would arrive at the following function for the energy per frame of inference (or equivalently for an LLM, Joules/Token): 

\begin{eqnarray}
E_{\text{frame}} &=& n_\text{weights}\ \ b_w\epsilon_{wM} \nonumber \\
& &+ T\ \ \big(n_\text{mac}(b_a\epsilon_{aM} + E_\text{synapse}) \nonumber \\
& &+ \ \ n_{\text{neurons}_1}(E_{\text{neuron}_1} +b_a\epsilon_{aM}) + \cdots  \nonumber \\
&&+ \ \ n_{\text{neurons}_N}(E_{\text{neuron}_N}+b_a\epsilon_{aM}) \big)
\label{eq:energy_full_inference}
\end{eqnarray}
where there may be N different types of nonlinearities in the model. One may want to test a model where different activations are stored in different memories, i.e $\epsilon_{aM}$ could be unique to each neuron type. This could be applied to the KV cache for instance, which might be larger or further from compute than other activation buffers. The granularity with which an estimation can be made can be adjusted as desired.

\begin{table*}
\centering
\begin{ruledtabular}
\begin{tabular}{@{}lll@{}}
Symbol & Description & Quantities for Fig.\ref{fig:energy_dnn}\\
\hline
$N$ & Number of neurons & \\
$n$ & Fan-in to processing element & \\
$T$ & Number of samples & \\
$b_w$ & Bits per weight & 32 \\
$b_a$ & Bits per activation & 32,1\\
$\epsilon_{wM}$ & Energy to access 1b from weight memory & 3.97 $\frac{\text{pJ}}{\text{bit}}$\\
 $\epsilon_{aM}$& Energy to access 1b from activation memory&0.0384$\frac{\text{pJ}}{\text{bit}}$\\
$\epsilon_\text{mul}$ & Energy to multiply $b_w$ and $b_a$ & 8.989 pJ\\
$\epsilon_\text{add}$ & Energy to add $b_w + b_a$ numbers & 0.239 pJ\\
$\epsilon_\text{act}$ & Energy for activation & 0.087 pJ\\
$\epsilon_\text{rng}$ & Energy for RNG & 0.181 pJ\\
$\epsilon_\text{compare}$ & Energy for RNG-activation comparison & 0.022 pJ\\
$n_{\text{weights}}$ & Number of weights & 2202122\\
$n_{\text{compute}}$ & Number of compute operations amounting to MACs or AC & 12659946\\
$n_{\text{activations}}$ & Number of activations & 118794\\
\end{tabular}
\end{ruledtabular}
\caption{Variables for p-DNN analysis}
\label{tab:energy_eq_variables}
\end{table*}

\section{Comparing FPGA implementations of DNN and p-DNN} \label{sec:fpga}
\begin{figure}
    \includegraphics[width=1\linewidth]{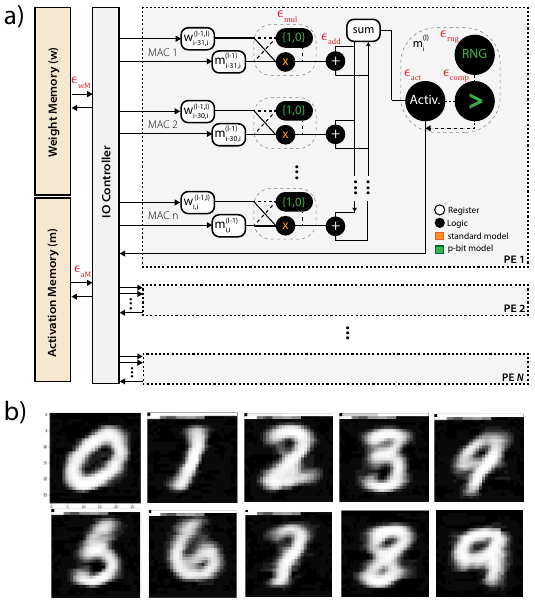} 
    \caption{(a) The processing element architecture for a building block, as implemented on the FPGA. (b) Image samples generated by the fpga with 1 probabilistic sample.}
    \label{fig:processing_element}
\end{figure}
We implement a both a P-DNN with p-bit activations and a traditional DNN with sigmoid activations in FPGA to validate the energy framework. FPGA based hardware implementations of probabilistic models outside of stochastic computing approaches have dealt exclusively with  energy based models \cite{pervaiz_weighted_2019, sutton_autonomous_2020, singh_hardware_2023}. On the other hand, this is not an energy based application, though the general architecture is intentionally kept very similar to existing p-bit hardware solutions, as seen in Fig.\ref{fig:processing_element}. In keeping the design similar to existing p-bit hardware solutions, a few key decisions were taken: 
\begin{itemize}
    \item Only fully connected layers are implemented, as opposed to convolutional layers that may require sliding windows. 
    \item The fan-in to any neuron in the network is limited to 32, similar to how e.g. k-nearest neighbor designs for Boltzmann machines limit fan-in to k. 
\end{itemize}

The image generation task is implemented via a DNN based on a Variational Autoencoder, shown in Fig.\ref{fig:model-architectures}c, trained on the MNIST dataset. This DNN is mapped onto a hardware processing element as described by Fig.\ref{fig:processing_element}a. Two versions of this DNN are trained, one that uses p-bits and one that uses standard 32b $\sigma$ activations, and consequently two versions of the hardware are built. There are 2 differences between the designs. First, the \textit{synapse} module implements a MAC operation for the standard design as opposed to AC for the p-bit design. Second, the \textit{neuron} module skips the RNG and comparison operations for the standard design. These changes are summarized in Fig.~\ref{fig:processing_element}.

\begin{table}[ht]
\centering
\begin{ruledtabular}

\begin{tabular}{@{}lccccc@{}}
 & Power (W) & Time (ns) & Energy (nJ) & LUTs & Regs \\
\hline
\multicolumn{6}{l}{\textbf{Full Design}} \\
p-bit  & 0.066  & 2.04e6 & 1.35e5  & 3796 & 2926 \\
tanh   & 0.168  & 2.04e6 & 3.43e5  & 4158 & 2759 \\
\hline
\multicolumn{6}{l}{\textbf{LFSR (RNG)}} \\
p-bit  & 1.4e-4 & --     & --      & 16   & 33 \\
\hline
\multicolumn{6}{l}{\textbf{Compare}} \\
p-bit  & 3.3e-4 & --     & --      & 17   & 33 \\
\end{tabular}
\end{ruledtabular}
\caption{Energy and utilization for FPGA designs on the ZCU104.}
\label{tab:performance}
\end{table}

\begin{figure*}
    \includegraphics[width=1\linewidth]{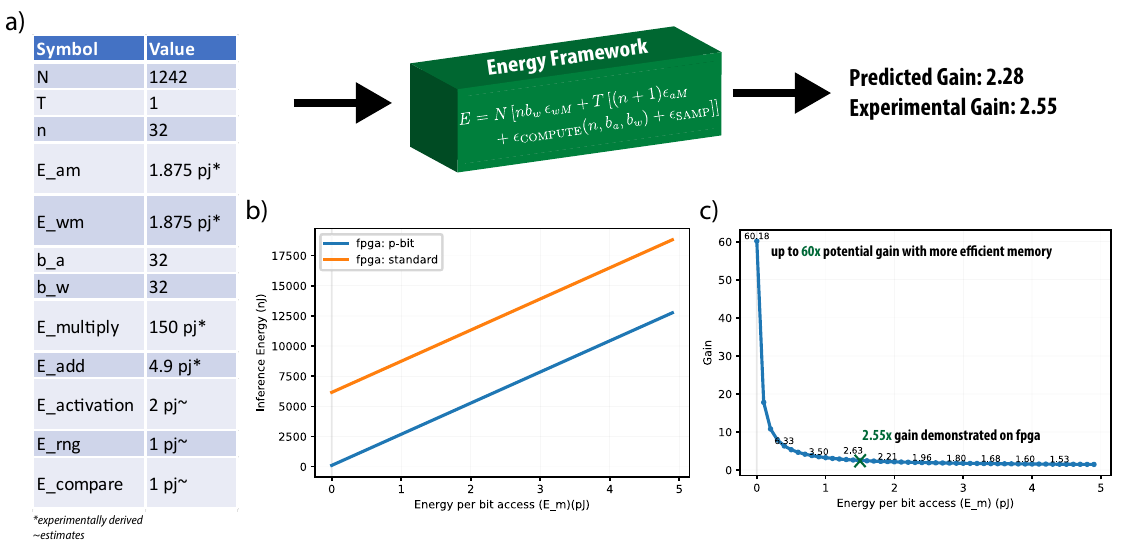} 
    \caption{(a) contains values used for the FPGA design's energy prediction. (b) is the predicted energy per frame of inference on the fpga. (c) shows the predicted gain curve comparing probabilistic and deterministic models across changing memory efficiency ($E_{wM} = E_{am} = E_M$)}
    \label{fig:fpga_prediction}
\end{figure*}

The power and energy of these designs are compared in Tab.\ref{tab:performance}. We observe a 4.42x improvement in the energy consumption of the mac module alone, which leads to an overall energy improvement of 2.55x. This aligns with the gain prediction in Fig.~\ref{fig:fpga_prediction}, where a 1.875 pJ/bit memory energy (approximate energy/bit for BRAM on the ZU7EV fpga) corresponds to 
2.3x gain, aligning closely with our experimentally obtained gain of 2.5x. With more efficient memory, this gain could be much larger as shown in Fig.~\ref{fig:fpga_prediction}.b and Fig.\ref{fig:fpga_prediction}.c, motivating the incorporation of in-memory-compute architectures. We also observe that mac-related CLB LUT usage goes down by 2.9x. 

A key consideration of using probabilistic models is the overhead that comes from the RNG over deterministic models. We see that the RNG and comparison overhead is negligible in terms of energy and area compared to the rest of the inference pipeline in custom hardware, which will only grow smaller as $n$ grows larger.

\section{Discussion}\label{sec6}
\textbf{Benefits over quantized model inference:} The sample-aware training algorithm outlined in Sec.\ref{sec3} paired with the energy analysis in Fig.\ref{fig:energy_dnn} provide some interesting insights. One possibility is the drawing of samples in parallel across separate hardware units, so long as each sample is independent and identically distributed (i.i.d), which they are for these models, and combining the samples at the output of the last layer. The internal math is all 1-bit, but the result could be the equivalent of 3-bit accuracy, delivered in less time. Such a hardware system could be a perfect fit for latency sensitive situations where energy and resources are not a concern. 

Another possibility for taking advantage of sampling is to build an $n$-bit p-DNN, which would match the energy of an $n$-bit DNN to within the difference in neuron energy ($\epsilon_\text{RNG} + \epsilon_\text{compare}$). Adding an additional sample to the inference pipeline could lead to 2+\% in accuracy, while increasing energy by just $0.7\%$ (see Fig.\ref{fig:energy_dnn}b) for models that run on DRAM. We consider our estimate conservative in that larger, production AI models, like LLMs, would use significantly more memory-related energy, making the additional sampling energy minuscule. 

An important advantage to doing inference via probabilistic systems over direct quantization is the ability to adjust the energy and accuracy at run-time through the number of samples. We show that a 1-bit probabilistic inference engine can be the equivalent of a 1-bit, 2-bit, and 3-bit quantized inference engine rolled into one. While each comparison on its own leads to roughly parity energy use, this is a giant practical leap, that a single hardware can serve all 3 bit-widths. \\

\textbf{An evolution on existing p-bit research:}
The FPGA experiment validates a couple of things. First, it supports the energy framework by closely matching predicted energy draw given the terms of the equation. Secondly, it shows that the same building block used for Boltzmann machines can implement p-DNNs. The idea is to show that p-DNNs can be implemented on near-BM hardware, for which there has already been extensive work done to build devices for and optimize at the system level \cite{aadit_massively_2022, sutton_autonomous_2020, daniel_experimental_2024}. 

How is it that machinery designed for Boltzmann machines can be applied to feed forward DNNs? Graph colored Ising models, for instance, enforce updating a subset of the total neurons per timestep such that every neuron in a model with $N_c$ colors would have an attempt to flip in $N_c$ timesteps. The neurons of each color could be thought of as a layer in a feed forward neural network; an MLP with p-bit activations could be seen as a rearranged graph-colored Ising model. In the same way Ising models with limited connectivity per neuron are able to be mapped to efficient in-memory-compute architectures \cite{chowdhury_accelerated_2023}, limited connectivity MLPs could be mapped to in-memory-computing architectures for ML tasks. The right balance between connectivity and performance could be application-specific. 

An important benefit to implementing p-bits in a feed forward fashion is that the backpropagation algorithm can be used to train the model, as opposed to standard energy-based training schemes like hebbian learning or contrastive divergence which are notably less efficient. \\

\textbf{Relevance to LLMs:}
Feed forward machine learning models are at the core of what makes AI so powerful today. While high bit weight and activation models (W32A32) were common a decade ago, the energy cost of running today's AI models has skyrocketed. The energy benefits of going to W$x$A$y$ models where $x$ and $y$ are small are clear, making it a highly active area of research.  Recent works have reduced weight precision in large language models (LLMs) to 4 bits and beyond \cite{noauthor_unslothdeepseek-v3-0324-gguf_2025, ma_era_2024}, but activation precision has remained challenging to reduce below 8 bits due to accuracy falloff. More so than weights, activations need to be adaptable at runtime, able to handle large scale outliers that may arise during inference \cite{xiao_smoothquant_2024, dettmers_llmint8_2022}. This is where p-bits may provide a means to scaling down LLM activation bits without compromising on accuracy. A 4-bit activation LLM for instance may achieve the required accuracy using a few samples. With fine-tuning and retraining methods like sample-aware training, that accuracy may be boosted, motivating research in that direction. \\

\textbf{Memory efficiency:}
We'd like to note that as it stands, the energy efficiency of p-DNNs, similar to standard DNNs, will depend more on memory technology compared to changes in other facets (tech-node, model architecture, etc). In order for p-bits to outperform standard quantization on a purely \textit{compute energy} front (synapse + neuron energy), advancements in sampling based training schemes and implementation that enable drawing fewer samples will be critical. One example of this is finding more effective ways to combine different samples, beyond simple averages. The flip side to this, as we noted previously, is that additional compute on the same data, in the vein of sampling, can improve accuracy without any notable impact on end-to-end energy.\\

\textbf{Related works: }Stochastic inference for feed forward CNNs was studied in \cite{li_binary-stochasticity-enabled_2024}but we were not able to replicate their results. While \cite{li_binary-stochasticity-enabled_2024} state that stochastic inference surpasses HP models within 25 samples for CIFAR10 classification, we find that accuracy comes to within 1\% but does not quite match. On the other hand, models implementing deterministic inference have found great success. For vision applications, W1A1 models have proven to be quite commendable \cite{courbariaux_binaryconnect_2016, hubara_quantized_nodate, courbariaux_binarized_2016}. Most recently, \cite{shi_expanding-and-shrinking_2025} have set state of the art results using W1A1 models for vision applications. 

\section{\label{sec7} Conclusion}
P-bits are a versatile tool with a proven track record in building highly energy efficient platforms for sampling and optimization problems, now applied to DNNs. We discuss two compute-only sampling schemes that can improve on the accuracy of standard DNNs with minor end-to-end energy penalties, demonstrated on image classification and generation. We pair this with an energy framework for probabilistic systems based on the basic building block consisting of a p-bit and its synapse. This building block is mapped to the p-bit based ASIC in \cite{li_122_2025} to reveal its energy components. We further support our framework with an FPGA demonstration, and show closely matching energy improvements to the framework's prediction. This manuscript presents foundational work for building energy efficient, sampling based DNNs. Future work could study analog implementations for p-bit based DNNs, integrating with in-memory compute architectures, or extending these results to beyond-vision applications. 

\section{Author Contributions}
LG set up the machine learning pipelines, conducted and recorded training and inference experiments, and built the FPGA design. ML and SS provided an energy breakdown of a QMC-ASIC, and conducted additional simulations to get energy estimates for core operations on a TSMC-65nm platform. All authors were critical in establishing the probabilistic methods discussed and demonstrated in the manuscript. LG and SD lead writing the paper, with contributions/revisions from all authors. All authors read and approved the final manuscript.

\section{Funding Declaration}
This work was partially supported by ONR-MURI Grant No. N000142312708, OptNet: Optimization with p-Bit Networks.

\section{Competing Interests}
Authors LG, SD, and BB have financial interests in Ludwig Computing. Author SS has a financial interest in Ixana.

\appendix
\section{Simulation Details \label{App:simulation_details}}
This appendix contains simulation details for training the DNNs used in this manuscript. Namely, this includes the classification model, and two image generation models, one of which was run on the FPGA. Code is available on request. 

For classification experiments, we went with sigmoid for the p-bit implementation to give deterministic models the benefit of the doubt, as sigmoid activations performed much better for deterministic models. For image generation, we found no notable difference in performance between sigmoid and tanh, and so stuck with tanh p-bits and tanh deterministic baselines. For each experiment, the derivative of the p-bit was implemented as the derivative of the activation, going "straight-through" the sign function \cite{shekhovtsov_reintroducing_2021}. All models were trained on an RTX 4070 super, and wandb was used to log and track experiments. 

\subsection{Classification Experiment} 
The classification experiment was carried out on the CIFAR10 dataset. This dataset contains (32,32,3) dimensional images that span 10 classes: plane, car, bird, cat, deer, dog, frog, horse, ship, truck. The data is split between 50000 training images and 10000 test images.

\begin{figure*}
    \includegraphics[width=1\linewidth]{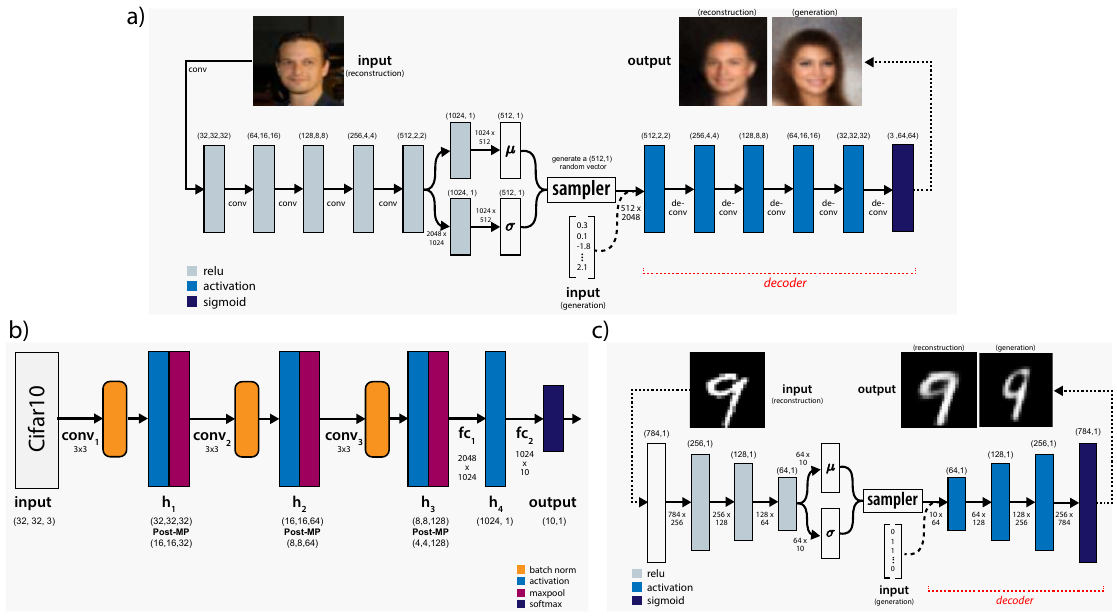} 
    \caption{(a) Model for celebrity image generation experiment. (b) Model for CIFAR10 classification experiment. (c) Model for MNIST generation FPGA experiment.}
    \label{fig:model-architectures}
\end{figure*}

The architecture is depicted in Fig.\ref{fig:model-architectures}. All models are trained for 1000 epochs using the ADAM optimizer on cross entropy loss using pytorch. All models use batch size of 64, have a constant learning rate of 1e-3. There is a standard horizontal flip pre-processing on the dataset during training for all models. Batch norm can be folded into the preceding convolutional or fully connected layer during inference \cite{yvinec_fold_2022}, allowing us to take advantage of it during training without any penalty on inference energy.


The CIFAR10-classification models are initialized with weights following Glorot (xavier) weight initialization from \cite{glorot_understanding_nodate}. Glorot initialization is a one-time overhead that ensure that the weights of the model have the same variance per layer despite changing layer configurations, and are centered around 0. It can be described as the following. 
\begin{equation}
W_{ij} \sim \mathcal{N}\!\Bigl(0,\;\tfrac{2}{\mathrm{fan\_in} + \mathrm{fan\_out}}\Bigr)
\end{equation}
where $i$ and $j$ are the input and output units of weight matrix $W$. Fan\_in describes the input size of weight layer $W$ and fan\_out describes the output size. Biases are set to 0. 

\subsection{Celebrity Generation Experiment}
The images shown in Fig. \ref{fig:image_gen} are from a variational autoencoder that was trained on the celeba dataset \cite{liu_deep_2015}. This dataset consists of 202,599 face images of dimension $178\times218\times3$, of which 162,770 images were used as the training set and 19,962 images were used as the test set (the remaining 19,867 validation images were not used). The images were scaled down to $64 \times 64 \times 3$ via torch.resize() to speed up training and make experimentation more feasible. 

The VAE uses 5 convolutional layers for the encoder (3, 32, 64, 128, 256, 512) with kernel size 4, stride 2, padding 1, and the same layers in reverse as transposed convolutions for the decoder. A latent dimension of 512 was used. Training used binary cross entropy loss paired with the ADAM optimizer. Training was carried out for 1000 epochs, using a learning rate of 1e-4 and a batch size of 64.  

\subsection{FPGA Generation Experiment}
A Variational Autoencoder (VAE) is trained to generate the MNIST digits following the architecture in Fig.\ref{fig:model-architectures}c. The deterministic and probabilistic versions of this design are implemented on a ZCU104 board, which houses a Zynq Ultrascale+ ZU7EV FPGA.  The weights for all models are converted to fixed point Q6.25. The p-bit design uses a single sample, which proved sufficient for generating the images. The design was run at 100 MHz. The input is deterministically binarized before feeding into the first layer of the model. The full VAE was trained using a GPU, and the decoder was implemented on the FPGA. For the MAC operation, we implement a high precision intermediate accumulate as is standard in FMAC units.

\section{Additional Experiments}

\subsection{Classification with tanh-p-bits}
\begin{figure}
    \includegraphics[width=1\linewidth]{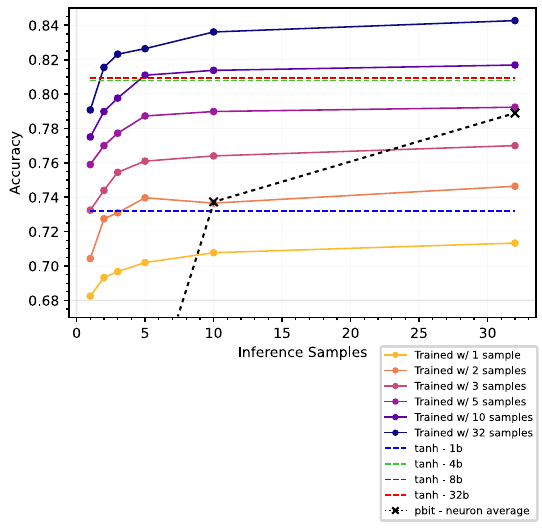} 
    \caption{Results for  classification using traditional tanh based p-bits, which observe a significant benefit over deterministic tanh activations compared to sigmoid.}
    \label{fig:tanh_accuracy}
\end{figure}

The same classification experiments were carried out using traditional p-bit activations. The probabilistic models seem to train roughly equivalently to those involving sigmoid activations, though the deterministic models do not. This suggests that probabilistic models have some resilience to training shortcomings that affect traditional tanh models, like the vanishing gradient problem. The architecture used is the same as shown in Fig.\ref{fig:model-architectures}, and the results are shown in Fig.\ref{fig:tanh_accuracy}. 

\begin{figure}
    \includegraphics[width=1\linewidth]{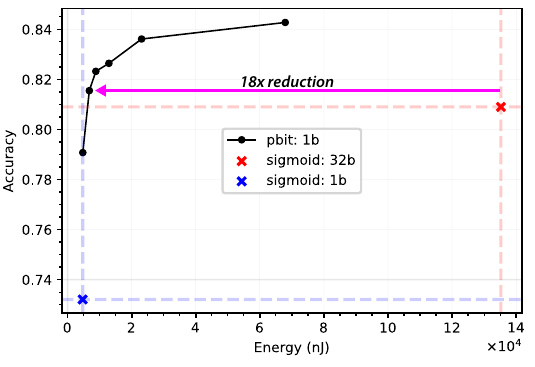} 
    \caption{Results for  classification using traditional tanh based p-bits, which observe a significant benefit over deterministic tanh activations compared to sigmoid.}
    \label{fig:tanh_energy}
\end{figure}

The energy verses accuracy curve for the tanh model would change since the accuracy per sample has changed relative to the deterministic baseline. However, energy per sample calculations would be unchanged, as only the LUT entries in $\epsilon_\text{act}$ are modified with those of a different nonlinearity. The updated energy vs accuracy curve is shown in Fig.\ref{fig:tanh_energy}. 

\subsection{Energy of tanh vs sigmoid p-bits}
Inside a PE that implements a \textit{sigmoid} based probabilistic bit, the PE only needs to sum weights paired to +1 activations, as the remaining weights are paired with 0s. For a PE that takes $n$ inputs, this results in $\approx 0.5n$ operations assuming roughly half the p-bits are in a +1 state. In the case of a \textit{tanh} based p-bit, activations are bipolar instead of binary. For $n$ weight inputs to a neuron, this results in $\approx1.5n$ operations where $n$ operations are attributed to summations, and $0.5n$ operations attributed to sign-flipping weights paired to -1 activations. 

However, by a simple reformulation, the tanh approach can also be reduced to $\approx 0.5n$ operations for any reasonably large $n$ input PE. 
$$
S = \sum_i w_im_i
$$
$$
S_{weights} = \sum_i w_i
$$
$$
S_1 = \sum_i w_im_i ~\forall~ m_i=1
$$
$$
S = S_{weights}-2(S_{weights} - S_1) 
$$
Let's say w = [1, 2, 3, 4]. m = [1, 1, -1, -1]. Then we have $S_{weights}=10$, $S_1=3$, which makes $S=-4$. This allows us to skip sign-flips, and only keep track of $S_{weights}$ in addition to the baseline sum of 1-paired weights, which can be loaded into the hardware as a pre-inference step. Though this adds $n_{activations}b_w$ memory overhead to the model, it need only be calculated once for a trained model for which inference can be done any number of times. 

During inference, this method increases total PE computation by 2 adds, and requires 1 additional memory read ($S_{weights}$). For larger models that have hundreds of inputs per neuron, this inference overhead becomes negligible. For that reason, we assume that probabilistic computation for both tanh and sigmoid based probabilistic bits require only summing weights paired with +1 activations. 


\newpage
\clearpage

\bibliography{apssamp}

\begin{thebibliography}{24}%
\makeatletter
\providecommand \@ifxundefined [1]{%
 \@ifx{#1\undefined}
}%
\providecommand \@ifnum [1]{%
 \ifnum #1\expandafter \@firstoftwo
 \else \expandafter \@secondoftwo
 \fi
}%
\providecommand \@ifx [1]{%
 \ifx #1\expandafter \@firstoftwo
 \else \expandafter \@secondoftwo
 \fi
}%
\providecommand \natexlab [1]{#1}%
\providecommand \enquote  [1]{``#1''}%
\providecommand \bibnamefont  [1]{#1}%
\providecommand \bibfnamefont [1]{#1}%
\providecommand \citenamefont [1]{#1}%
\providecommand \href@noop [0]{\@secondoftwo}%
\providecommand \href [0]{\begingroup \@sanitize@url \@href}%
\providecommand \@href[1]{\@@startlink{#1}\@@href}%
\providecommand \@@href[1]{\endgroup#1\@@endlink}%
\providecommand \@sanitize@url [0]{\catcode `\\12\catcode `\$12\catcode `\&12\catcode `\#12\catcode `\^12\catcode `\_12\catcode `\%12\relax}%
\providecommand \@@startlink[1]{}%
\providecommand \@@endlink[0]{}%
\providecommand \url  [0]{\begingroup\@sanitize@url \@url }%
\providecommand \@url [1]{\endgroup\@href {#1}{\urlprefix }}%
\providecommand \urlprefix  [0]{URL }%
\providecommand \Eprint [0]{\href }%
\providecommand \doibase [0]{https://doi.org/}%
\providecommand \selectlanguage [0]{\@gobble}%
\providecommand \bibinfo  [0]{\@secondoftwo}%
\providecommand \bibfield  [0]{\@secondoftwo}%
\providecommand \translation [1]{[#1]}%
\providecommand \BibitemOpen [0]{}%
\providecommand \bibitemStop [0]{}%
\providecommand \bibitemNoStop [0]{.\EOS\space}%
\providecommand \EOS [0]{\spacefactor3000\relax}%
\providecommand \BibitemShut  [1]{\csname bibitem#1\endcsname}%
\let\auto@bib@innerbib\@empty
\bibitem [{\citenamefont {Hubara}\ \emph {et~al.}()\citenamefont {Hubara}, \citenamefont {Courbariaux}, \citenamefont {Soudry}, \citenamefont {El-Yaniv},\ and\ \citenamefont {Bengio}}]{hubara_quantized_nodate}%
  \BibitemOpen
  \bibfield  {author} {\bibinfo {author} {\bibfnamefont {I.}~\bibnamefont {Hubara}}, \bibinfo {author} {\bibfnamefont {M.}~\bibnamefont {Courbariaux}}, \bibinfo {author} {\bibfnamefont {D.}~\bibnamefont {Soudry}}, \bibinfo {author} {\bibfnamefont {R.}~\bibnamefont {El-Yaniv}},\ and\ \bibinfo {author} {\bibfnamefont {Y.}~\bibnamefont {Bengio}},\ }\href@noop {} {\bibinfo {title} {Quantized {Neural} {Networks}: {Training} {Neural} {Networks} with {Low} {Precision} {Weights} and {Activations}}}\BibitemShut {NoStop}%
\bibitem [{\citenamefont {Raiko}\ \emph {et~al.}(2015)\citenamefont {Raiko}, \citenamefont {Berglund}, \citenamefont {Alain},\ and\ \citenamefont {Dinh}}]{raiko_techniques_2015}%
  \BibitemOpen
  \bibfield  {author} {\bibinfo {author} {\bibfnamefont {T.}~\bibnamefont {Raiko}}, \bibinfo {author} {\bibfnamefont {M.}~\bibnamefont {Berglund}}, \bibinfo {author} {\bibfnamefont {G.}~\bibnamefont {Alain}},\ and\ \bibinfo {author} {\bibfnamefont {L.}~\bibnamefont {Dinh}},\ }\href {https://doi.org/10.48550/arXiv.1406.2989} {\bibinfo {title} {Techniques for {Learning} {Binary} {Stochastic} {Feedforward} {Neural} {Networks}}} (\bibinfo {year} {2015}),\ \bibinfo {note} {arXiv:1406.2989}\BibitemShut {NoStop}%
\bibitem [{\citenamefont {Capra}\ \emph {et~al.}(2020)\citenamefont {Capra}, \citenamefont {Bussolino}, \citenamefont {Marchisio}, \citenamefont {Masera}, \citenamefont {Martina},\ and\ \citenamefont {Shafique}}]{capra_hardware_2020}%
  \BibitemOpen
  \bibfield  {author} {\bibinfo {author} {\bibfnamefont {M.}~\bibnamefont {Capra}}, \bibinfo {author} {\bibfnamefont {B.}~\bibnamefont {Bussolino}}, \bibinfo {author} {\bibfnamefont {A.}~\bibnamefont {Marchisio}}, \bibinfo {author} {\bibfnamefont {G.}~\bibnamefont {Masera}}, \bibinfo {author} {\bibfnamefont {M.}~\bibnamefont {Martina}},\ and\ \bibinfo {author} {\bibfnamefont {M.}~\bibnamefont {Shafique}},\ }\bibfield  {title} {\bibinfo {title} {Hardware and {Software} {Optimizations} for {Accelerating} {Deep} {Neural} {Networks}: {Survey} of {Current} {Trends}, {Challenges}, and the {Road} {Ahead}},\ }\href {https://doi.org/10.1109/ACCESS.2020.3039858} {\bibfield  {journal} {\bibinfo  {journal} {IEEE Access}\ }\textbf {\bibinfo {volume} {8}},\ \bibinfo {pages} {225134} (\bibinfo {year} {2020})}\BibitemShut {NoStop}%
\bibitem [{\citenamefont {Jacob}\ \emph {et~al.}(2018)\citenamefont {Jacob}, \citenamefont {Kligys}, \citenamefont {Chen}, \citenamefont {Zhu}, \citenamefont {Tang}, \citenamefont {Howard}, \citenamefont {Adam},\ and\ \citenamefont {Kalenichenko}}]{jacob_quantization_2018}%
  \BibitemOpen
  \bibfield  {author} {\bibinfo {author} {\bibfnamefont {B.}~\bibnamefont {Jacob}}, \bibinfo {author} {\bibfnamefont {S.}~\bibnamefont {Kligys}}, \bibinfo {author} {\bibfnamefont {B.}~\bibnamefont {Chen}}, \bibinfo {author} {\bibfnamefont {M.}~\bibnamefont {Zhu}}, \bibinfo {author} {\bibfnamefont {M.}~\bibnamefont {Tang}}, \bibinfo {author} {\bibfnamefont {A.}~\bibnamefont {Howard}}, \bibinfo {author} {\bibfnamefont {H.}~\bibnamefont {Adam}},\ and\ \bibinfo {author} {\bibfnamefont {D.}~\bibnamefont {Kalenichenko}},\ }\bibfield  {title} {\bibinfo {title} {Quantization and {Training} of {Neural} {Networks} for {Efficient} {Integer}-{Arithmetic}-{Only} {Inference}},\ }in\ \href {https://doi.org/10.1109/CVPR.2018.00286} {\emph {\bibinfo {booktitle} {2018 {IEEE}/{CVF} {Conference} on {Computer} {Vision} and {Pattern} {Recognition}}}}\ (\bibinfo {year} {2018})\ pp.\ \bibinfo {pages} {2704--2713},\ \bibinfo {note} {iSSN: 2575-7075}\BibitemShut {NoStop}%
\bibitem [{\citenamefont {Li}\ \emph {et~al.}(2025)\citenamefont {Li}, \citenamefont {Ghosh}, \citenamefont {Jaiswal}, \citenamefont {Ghantasala}, \citenamefont {Behin-Aein}, \citenamefont {Sen},\ and\ \citenamefont {Datta}}]{li_122_2025}%
  \BibitemOpen
  \bibfield  {author} {\bibinfo {author} {\bibfnamefont {M.-C.}\ \bibnamefont {Li}}, \bibinfo {author} {\bibfnamefont {A.}~\bibnamefont {Ghosh}}, \bibinfo {author} {\bibfnamefont {R.}~\bibnamefont {Jaiswal}}, \bibinfo {author} {\bibfnamefont {L.~A.}\ \bibnamefont {Ghantasala}}, \bibinfo {author} {\bibfnamefont {B.}~\bibnamefont {Behin-Aein}}, \bibinfo {author} {\bibfnamefont {S.}~\bibnamefont {Sen}},\ and\ \bibinfo {author} {\bibfnamefont {S.}~\bibnamefont {Datta}},\ }\bibfield  {title} {\bibinfo {title} {12.2 p-{Circuits}: {Neither} {Digital} {Nor} {Analog}},\ }in\ \href {https://doi.org/10.1109/ISSCC49661.2025.10904553} {\emph {\bibinfo {booktitle} {2025 {IEEE} {International} {Solid}-{State} {Circuits} {Conference} ({ISSCC})}}},\ Vol.~\bibinfo {volume} {68}\ (\bibinfo {year} {2025})\ pp.\ \bibinfo {pages} {1--3},\ \bibinfo {note} {iSSN: 2376-8606}\BibitemShut {NoStop}%
\bibitem [{\citenamefont {O'Connor}\ \emph {et~al.}(2017)\citenamefont {O'Connor}, \citenamefont {Chatterjee}, \citenamefont {Lee}, \citenamefont {Wilson}, \citenamefont {Agrawal}, \citenamefont {Keckler},\ and\ \citenamefont {Dally}}]{oconnor_fine-grained_2017}%
  \BibitemOpen
  \bibfield  {author} {\bibinfo {author} {\bibfnamefont {M.}~\bibnamefont {O'Connor}}, \bibinfo {author} {\bibfnamefont {N.}~\bibnamefont {Chatterjee}}, \bibinfo {author} {\bibfnamefont {D.}~\bibnamefont {Lee}}, \bibinfo {author} {\bibfnamefont {J.}~\bibnamefont {Wilson}}, \bibinfo {author} {\bibfnamefont {A.}~\bibnamefont {Agrawal}}, \bibinfo {author} {\bibfnamefont {S.~W.}\ \bibnamefont {Keckler}},\ and\ \bibinfo {author} {\bibfnamefont {W.~J.}\ \bibnamefont {Dally}},\ }\bibfield  {title} {\bibinfo {title} {Fine-grained {DRAM}: energy-efficient {DRAM} for extreme bandwidth systems},\ }in\ \href {https://doi.org/10.1145/3123939.3124545} {\emph {\bibinfo {booktitle} {Proceedings of the 50th {Annual} {IEEE}/{ACM} {International} {Symposium} on {Microarchitecture}}}}\ (\bibinfo  {publisher} {ACM},\ \bibinfo {address} {Cambridge Massachusetts},\ \bibinfo {year} {2017})\ pp.\ \bibinfo {pages} {41--54}\BibitemShut {NoStop}%
\bibitem [{\citenamefont {Pervaiz}\ \emph {et~al.}(2019)\citenamefont {Pervaiz}, \citenamefont {Sutton}, \citenamefont {Ghantasala},\ and\ \citenamefont {Camsari}}]{pervaiz_weighted_2019}%
  \BibitemOpen
  \bibfield  {author} {\bibinfo {author} {\bibfnamefont {A.~Z.}\ \bibnamefont {Pervaiz}}, \bibinfo {author} {\bibfnamefont {B.~M.}\ \bibnamefont {Sutton}}, \bibinfo {author} {\bibfnamefont {L.~A.}\ \bibnamefont {Ghantasala}},\ and\ \bibinfo {author} {\bibfnamefont {K.~Y.}\ \bibnamefont {Camsari}},\ }\bibfield  {title} {\bibinfo {title} {Weighted p -{Bits} for {FPGA} {Implementation} of {Probabilistic} {Circuits}},\ }\href {https://doi.org/10.1109/TNNLS.2018.2874565} {\bibfield  {journal} {\bibinfo  {journal} {IEEE Transactions on Neural Networks and Learning Systems}\ }\textbf {\bibinfo {volume} {30}},\ \bibinfo {pages} {1920} (\bibinfo {year} {2019})},\ \bibinfo {note} {conference Name: IEEE Transactions on Neural Networks and Learning Systems}\BibitemShut {NoStop}%
\bibitem [{\citenamefont {Sutton}\ \emph {et~al.}(2020)\citenamefont {Sutton}, \citenamefont {Faria}, \citenamefont {Ghantasala}, \citenamefont {Jaiswal}, \citenamefont {Camsari},\ and\ \citenamefont {Datta}}]{sutton_autonomous_2020}%
  \BibitemOpen
  \bibfield  {author} {\bibinfo {author} {\bibfnamefont {B.}~\bibnamefont {Sutton}}, \bibinfo {author} {\bibfnamefont {R.}~\bibnamefont {Faria}}, \bibinfo {author} {\bibfnamefont {L.~A.}\ \bibnamefont {Ghantasala}}, \bibinfo {author} {\bibfnamefont {R.}~\bibnamefont {Jaiswal}}, \bibinfo {author} {\bibfnamefont {K.~Y.}\ \bibnamefont {Camsari}},\ and\ \bibinfo {author} {\bibfnamefont {S.}~\bibnamefont {Datta}},\ }\bibfield  {title} {\bibinfo {title} {Autonomous {Probabilistic} {Coprocessing} {With} {Petaflips} per {Second}},\ }\href {https://doi.org/10.1109/ACCESS.2020.3018682} {\bibfield  {journal} {\bibinfo  {journal} {IEEE Access}\ }\textbf {\bibinfo {volume} {8}},\ \bibinfo {pages} {157238} (\bibinfo {year} {2020})},\ \bibinfo {note} {29 citations (Semantic Scholar/DOI) [2023-05-30] Conference Name: IEEE Access}\BibitemShut {NoStop}%
\bibitem [{\citenamefont {Singh}\ \emph {et~al.}(2023)\citenamefont {Singh}, \citenamefont {Niazi}, \citenamefont {Chowdhury}, \citenamefont {Selcuk}, \citenamefont {Kaneko}, \citenamefont {Kobayashi}, \citenamefont {Kanai}, \citenamefont {Ohno}, \citenamefont {Fukami},\ and\ \citenamefont {Camsari}}]{singh_hardware_2023}%
  \BibitemOpen
  \bibfield  {author} {\bibinfo {author} {\bibfnamefont {N.~S.}\ \bibnamefont {Singh}}, \bibinfo {author} {\bibfnamefont {S.}~\bibnamefont {Niazi}}, \bibinfo {author} {\bibfnamefont {S.}~\bibnamefont {Chowdhury}}, \bibinfo {author} {\bibfnamefont {K.}~\bibnamefont {Selcuk}}, \bibinfo {author} {\bibfnamefont {H.}~\bibnamefont {Kaneko}}, \bibinfo {author} {\bibfnamefont {K.}~\bibnamefont {Kobayashi}}, \bibinfo {author} {\bibfnamefont {S.}~\bibnamefont {Kanai}}, \bibinfo {author} {\bibfnamefont {H.}~\bibnamefont {Ohno}}, \bibinfo {author} {\bibfnamefont {S.}~\bibnamefont {Fukami}},\ and\ \bibinfo {author} {\bibfnamefont {K.~Y.}\ \bibnamefont {Camsari}},\ }\bibfield  {title} {\bibinfo {title} {Hardware {Demonstration} of {Feedforward} {Stochastic} {Neural} {Networks} with {Fast} {MTJ}-based p-bits},\ }in\ \href {https://doi.org/10.1109/IEDM45741.2023.10413686} {\emph {\bibinfo {booktitle} {2023 {International} {Electron} {Devices} {Meeting} ({IEDM})}}}\ (\bibinfo {year} {2023})\ pp.\ \bibinfo {pages} {1--4},\
  \bibinfo {note} {iSSN: 2156-017X}\BibitemShut {NoStop}%
\bibitem [{\citenamefont {Aadit}\ \emph {et~al.}(2022)\citenamefont {Aadit}, \citenamefont {Grimaldi}, \citenamefont {Carpentieri}, \citenamefont {Theogarajan}, \citenamefont {Martinis}, \citenamefont {Finocchio},\ and\ \citenamefont {Camsari}}]{aadit_massively_2022}%
  \BibitemOpen
  \bibfield  {author} {\bibinfo {author} {\bibfnamefont {N.~A.}\ \bibnamefont {Aadit}}, \bibinfo {author} {\bibfnamefont {A.}~\bibnamefont {Grimaldi}}, \bibinfo {author} {\bibfnamefont {M.}~\bibnamefont {Carpentieri}}, \bibinfo {author} {\bibfnamefont {L.}~\bibnamefont {Theogarajan}}, \bibinfo {author} {\bibfnamefont {J.~M.}\ \bibnamefont {Martinis}}, \bibinfo {author} {\bibfnamefont {G.}~\bibnamefont {Finocchio}},\ and\ \bibinfo {author} {\bibfnamefont {K.~Y.}\ \bibnamefont {Camsari}},\ }\bibfield  {title} {\bibinfo {title} {Massively {Parallel} {Probabilistic} {Computing} with {Sparse} {Ising} {Machines}},\ }\bibfield  {journal} {\bibinfo  {journal} {Nature Electronics}\ }\href {https://doi.org/10.1038/s41928-022-00774-2} {10.1038/s41928-022-00774-2} (\bibinfo {year} {2022}),\ \bibinfo {note} {6 citations (Semantic Scholar/arXiv) [2022-07-08] 6 citations (Semantic Scholar/DOI) [2022-07-08] arXiv:2110.02481 [cond-mat]}\BibitemShut {NoStop}%
\bibitem [{\citenamefont {Daniel}\ \emph {et~al.}(2024)\citenamefont {Daniel}, \citenamefont {Sun}, \citenamefont {Zhang}, \citenamefont {Tan}, \citenamefont {Dilley}, \citenamefont {Chen},\ and\ \citenamefont {Appenzeller}}]{daniel_experimental_2024}%
  \BibitemOpen
  \bibfield  {author} {\bibinfo {author} {\bibfnamefont {J.}~\bibnamefont {Daniel}}, \bibinfo {author} {\bibfnamefont {Z.}~\bibnamefont {Sun}}, \bibinfo {author} {\bibfnamefont {X.}~\bibnamefont {Zhang}}, \bibinfo {author} {\bibfnamefont {Y.}~\bibnamefont {Tan}}, \bibinfo {author} {\bibfnamefont {N.}~\bibnamefont {Dilley}}, \bibinfo {author} {\bibfnamefont {Z.}~\bibnamefont {Chen}},\ and\ \bibinfo {author} {\bibfnamefont {J.}~\bibnamefont {Appenzeller}},\ }\bibfield  {title} {\bibinfo {title} {Experimental demonstration of an on-chip p-bit core based on stochastic magnetic tunnel junctions and {2D} {MoS2} transistors},\ }\href {https://doi.org/10.1038/s41467-024-48152-0} {\bibfield  {journal} {\bibinfo  {journal} {Nature Communications}\ }\textbf {\bibinfo {volume} {15}},\ \bibinfo {pages} {4098} (\bibinfo {year} {2024})},\ \bibinfo {note} {publisher: Nature Publishing Group}\BibitemShut {NoStop}%
\bibitem [{\citenamefont {Chowdhury}\ \emph {et~al.}(2023)\citenamefont {Chowdhury}, \citenamefont {Camsari},\ and\ \citenamefont {Datta}}]{chowdhury_accelerated_2023}%
  \BibitemOpen
  \bibfield  {author} {\bibinfo {author} {\bibfnamefont {S.}~\bibnamefont {Chowdhury}}, \bibinfo {author} {\bibfnamefont {K.~Y.}\ \bibnamefont {Camsari}},\ and\ \bibinfo {author} {\bibfnamefont {S.}~\bibnamefont {Datta}},\ }\bibfield  {title} {\bibinfo {title} {Accelerated quantum {Monte} {Carlo} with probabilistic computers},\ }\href {https://doi.org/10.1038/s42005-023-01202-3} {\bibfield  {journal} {\bibinfo  {journal} {Communications Physics}\ }\textbf {\bibinfo {volume} {6}},\ \bibinfo {pages} {1} (\bibinfo {year} {2023})},\ \bibinfo {note} {1 citations (Semantic Scholar/DOI) [2023-06-05] Number: 1 Publisher: Nature Publishing Group}\BibitemShut {NoStop}%
\bibitem [{noa(2025)}]{noauthor_unslothdeepseek-v3-0324-gguf_2025}%
  \BibitemOpen
  \href {https://huggingface.co/unsloth/DeepSeek-V3-0324-GGUF} {\bibinfo {title} {unsloth/{DeepSeek}-{V3}-0324-{GGUF} · {Hugging} {Face}}} (\bibinfo {year} {2025})\BibitemShut {NoStop}%
\bibitem [{\citenamefont {Ma}\ \emph {et~al.}(2024)\citenamefont {Ma}, \citenamefont {Wang}, \citenamefont {Ma}, \citenamefont {Wang}, \citenamefont {Wang}, \citenamefont {Huang}, \citenamefont {Dong}, \citenamefont {Wang}, \citenamefont {Xue},\ and\ \citenamefont {Wei}}]{ma_era_2024}%
  \BibitemOpen
  \bibfield  {author} {\bibinfo {author} {\bibfnamefont {S.}~\bibnamefont {Ma}}, \bibinfo {author} {\bibfnamefont {H.}~\bibnamefont {Wang}}, \bibinfo {author} {\bibfnamefont {L.}~\bibnamefont {Ma}}, \bibinfo {author} {\bibfnamefont {L.}~\bibnamefont {Wang}}, \bibinfo {author} {\bibfnamefont {W.}~\bibnamefont {Wang}}, \bibinfo {author} {\bibfnamefont {S.}~\bibnamefont {Huang}}, \bibinfo {author} {\bibfnamefont {L.}~\bibnamefont {Dong}}, \bibinfo {author} {\bibfnamefont {R.}~\bibnamefont {Wang}}, \bibinfo {author} {\bibfnamefont {J.}~\bibnamefont {Xue}},\ and\ \bibinfo {author} {\bibfnamefont {F.}~\bibnamefont {Wei}},\ }\href {http://arxiv.org/abs/2402.17764} {\bibinfo {title} {The {Era} of 1-bit {LLMs}: {All} {Large} {Language} {Models} are in 1.58 {Bits}}} (\bibinfo {year} {2024}),\ \bibinfo {note} {arXiv:2402.17764 [cs]}\BibitemShut {NoStop}%
\bibitem [{\citenamefont {Xiao}\ \emph {et~al.}(2024)\citenamefont {Xiao}, \citenamefont {Lin}, \citenamefont {Seznec}, \citenamefont {Wu}, \citenamefont {Demouth},\ and\ \citenamefont {Han}}]{xiao_smoothquant_2024}%
  \BibitemOpen
  \bibfield  {author} {\bibinfo {author} {\bibfnamefont {G.}~\bibnamefont {Xiao}}, \bibinfo {author} {\bibfnamefont {J.}~\bibnamefont {Lin}}, \bibinfo {author} {\bibfnamefont {M.}~\bibnamefont {Seznec}}, \bibinfo {author} {\bibfnamefont {H.}~\bibnamefont {Wu}}, \bibinfo {author} {\bibfnamefont {J.}~\bibnamefont {Demouth}},\ and\ \bibinfo {author} {\bibfnamefont {S.}~\bibnamefont {Han}},\ }\href {https://doi.org/10.48550/arXiv.2211.10438} {\bibinfo {title} {{SmoothQuant}: {Accurate} and {Efficient} {Post}-{Training} {Quantization} for {Large} {Language} {Models}}} (\bibinfo {year} {2024}),\ \bibinfo {note} {arXiv:2211.10438 [cs]}\BibitemShut {NoStop}%
\bibitem [{\citenamefont {Dettmers}\ \emph {et~al.}(2022)\citenamefont {Dettmers}, \citenamefont {Lewis}, \citenamefont {Belkada},\ and\ \citenamefont {Zettlemoyer}}]{dettmers_llmint8_2022}%
  \BibitemOpen
  \bibfield  {author} {\bibinfo {author} {\bibfnamefont {T.}~\bibnamefont {Dettmers}}, \bibinfo {author} {\bibfnamefont {M.}~\bibnamefont {Lewis}}, \bibinfo {author} {\bibfnamefont {Y.}~\bibnamefont {Belkada}},\ and\ \bibinfo {author} {\bibfnamefont {L.}~\bibnamefont {Zettlemoyer}},\ }\href {https://doi.org/10.48550/arXiv.2208.07339} {\bibinfo {title} {{LLM}.int8(): 8-bit {Matrix} {Multiplication} for {Transformers} at {Scale}}} (\bibinfo {year} {2022}),\ \bibinfo {note} {arXiv:2208.07339 [cs]}\BibitemShut {NoStop}%
\bibitem [{\citenamefont {Li}\ \emph {et~al.}(2024)\citenamefont {Li}, \citenamefont {Wang}, \citenamefont {Wang}, \citenamefont {Dou}, \citenamefont {Ma}, \citenamefont {Zhou}, \citenamefont {Zhang}, \citenamefont {Lepri}, \citenamefont {Zhang}, \citenamefont {Luo}, \citenamefont {Xu}, \citenamefont {Yang}, \citenamefont {Zhang}, \citenamefont {Li}, \citenamefont {Ielmini},\ and\ \citenamefont {Liu}}]{li_binary-stochasticity-enabled_2024}%
  \BibitemOpen
  \bibfield  {author} {\bibinfo {author} {\bibfnamefont {Y.}~\bibnamefont {Li}}, \bibinfo {author} {\bibfnamefont {W.}~\bibnamefont {Wang}}, \bibinfo {author} {\bibfnamefont {M.}~\bibnamefont {Wang}}, \bibinfo {author} {\bibfnamefont {C.}~\bibnamefont {Dou}}, \bibinfo {author} {\bibfnamefont {Z.}~\bibnamefont {Ma}}, \bibinfo {author} {\bibfnamefont {H.}~\bibnamefont {Zhou}}, \bibinfo {author} {\bibfnamefont {P.}~\bibnamefont {Zhang}}, \bibinfo {author} {\bibfnamefont {N.}~\bibnamefont {Lepri}}, \bibinfo {author} {\bibfnamefont {X.}~\bibnamefont {Zhang}}, \bibinfo {author} {\bibfnamefont {Q.}~\bibnamefont {Luo}}, \bibinfo {author} {\bibfnamefont {X.}~\bibnamefont {Xu}}, \bibinfo {author} {\bibfnamefont {G.}~\bibnamefont {Yang}}, \bibinfo {author} {\bibfnamefont {F.}~\bibnamefont {Zhang}}, \bibinfo {author} {\bibfnamefont {L.}~\bibnamefont {Li}}, \bibinfo {author} {\bibfnamefont {D.}~\bibnamefont {Ielmini}},\ and\ \bibinfo {author} {\bibfnamefont {M.}~\bibnamefont {Liu}},\ }\bibfield  {title} {\bibinfo {title}
  {Binary-{Stochasticity}-{Enabled} {Highly} {Efficient} {Neuromorphic} {Deep} {Learning} {Achieves} {Better}-than-{Software} {Accuracy}},\ }\href {https://doi.org/10.1002/aisy.202300399} {\bibfield  {journal} {\bibinfo  {journal} {Advanced Intelligent Systems}\ }\textbf {\bibinfo {volume} {6}},\ \bibinfo {pages} {2300399} (\bibinfo {year} {2024})},\ \bibinfo {note} {\_eprint: https://onlinelibrary.wiley.com/doi/pdf/10.1002/aisy.202300399}\BibitemShut {NoStop}%
\bibitem [{\citenamefont {Courbariaux}\ \emph {et~al.}(2016{\natexlab{a}})\citenamefont {Courbariaux}, \citenamefont {Bengio},\ and\ \citenamefont {David}}]{courbariaux_binaryconnect_2016}%
  \BibitemOpen
  \bibfield  {author} {\bibinfo {author} {\bibfnamefont {M.}~\bibnamefont {Courbariaux}}, \bibinfo {author} {\bibfnamefont {Y.}~\bibnamefont {Bengio}},\ and\ \bibinfo {author} {\bibfnamefont {J.-P.}\ \bibnamefont {David}},\ }\href {https://doi.org/10.48550/arXiv.1511.00363} {\bibinfo {title} {{BinaryConnect}: {Training} {Deep} {Neural} {Networks} with binary weights during propagations}} (\bibinfo {year} {2016}{\natexlab{a}}),\ \bibinfo {note} {arXiv:1511.00363 [cs]}\BibitemShut {NoStop}%
\bibitem [{\citenamefont {Courbariaux}\ \emph {et~al.}(2016{\natexlab{b}})\citenamefont {Courbariaux}, \citenamefont {Hubara}, \citenamefont {Soudry}, \citenamefont {El-Yaniv},\ and\ \citenamefont {Bengio}}]{courbariaux_binarized_2016}%
  \BibitemOpen
  \bibfield  {author} {\bibinfo {author} {\bibfnamefont {M.}~\bibnamefont {Courbariaux}}, \bibinfo {author} {\bibfnamefont {I.}~\bibnamefont {Hubara}}, \bibinfo {author} {\bibfnamefont {D.}~\bibnamefont {Soudry}}, \bibinfo {author} {\bibfnamefont {R.}~\bibnamefont {El-Yaniv}},\ and\ \bibinfo {author} {\bibfnamefont {Y.}~\bibnamefont {Bengio}},\ }\href {https://doi.org/10.48550/arXiv.1602.02830} {\bibinfo {title} {Binarized {Neural} {Networks}: {Training} {Deep} {Neural} {Networks} with {Weights} and {Activations} {Constrained} to +1 or -1}} (\bibinfo {year} {2016}{\natexlab{b}}),\ \bibinfo {note} {arXiv:1602.02830 [cs]}\BibitemShut {NoStop}%
\bibitem [{\citenamefont {Shi}\ \emph {et~al.}(2025)\citenamefont {Shi}, \citenamefont {Sun}, \citenamefont {Qi}, \citenamefont {Hao},\ and\ \citenamefont {Yang}}]{shi_expanding-and-shrinking_2025}%
  \BibitemOpen
  \bibfield  {author} {\bibinfo {author} {\bibfnamefont {X.}~\bibnamefont {Shi}}, \bibinfo {author} {\bibfnamefont {C.}~\bibnamefont {Sun}}, \bibinfo {author} {\bibfnamefont {Z.}~\bibnamefont {Qi}}, \bibinfo {author} {\bibfnamefont {L.}~\bibnamefont {Hao}},\ and\ \bibinfo {author} {\bibfnamefont {X.}~\bibnamefont {Yang}},\ }\href {https://doi.org/10.48550/arXiv.2503.23709} {\bibinfo {title} {Expanding-and-{Shrinking} {Binary} {Neural} {Networks}}} (\bibinfo {year} {2025}),\ \bibinfo {note} {arXiv:2503.23709 [cs]}\BibitemShut {NoStop}%
\bibitem [{\citenamefont {Shekhovtsov}\ and\ \citenamefont {Yanush}(2021)}]{shekhovtsov_reintroducing_2021}%
  \BibitemOpen
  \bibfield  {author} {\bibinfo {author} {\bibfnamefont {A.}~\bibnamefont {Shekhovtsov}}\ and\ \bibinfo {author} {\bibfnamefont {V.}~\bibnamefont {Yanush}},\ }\href {https://doi.org/10.48550/arXiv.2006.06880} {\bibinfo {title} {Reintroducing {Straight}-{Through} {Estimators} as {Principled} {Methods} for {Stochastic} {Binary} {Networks}}} (\bibinfo {year} {2021}),\ \bibinfo {note} {arXiv:2006.06880 [stat]}\BibitemShut {NoStop}%
\bibitem [{\citenamefont {Yvinec}\ \emph {et~al.}(2022)\citenamefont {Yvinec}, \citenamefont {Dapogny},\ and\ \citenamefont {Bailly}}]{yvinec_fold_2022}%
  \BibitemOpen
  \bibfield  {author} {\bibinfo {author} {\bibfnamefont {E.}~\bibnamefont {Yvinec}}, \bibinfo {author} {\bibfnamefont {A.}~\bibnamefont {Dapogny}},\ and\ \bibinfo {author} {\bibfnamefont {K.}~\bibnamefont {Bailly}},\ }\href {https://doi.org/10.48550/arXiv.2203.14646} {\bibinfo {title} {To {Fold} or {Not} to {Fold}: a {Necessary} and {Sufficient} {Condition} on {Batch}-{Normalization} {Layers} {Folding}}} (\bibinfo {year} {2022}),\ \bibinfo {note} {arXiv:2203.14646 [cs]}\BibitemShut {NoStop}%
\bibitem [{\citenamefont {Glorot}\ and\ \citenamefont {Bengio}()}]{glorot_understanding_nodate}%
  \BibitemOpen
  \bibfield  {author} {\bibinfo {author} {\bibfnamefont {X.}~\bibnamefont {Glorot}}\ and\ \bibinfo {author} {\bibfnamefont {Y.}~\bibnamefont {Bengio}},\ }\href@noop {} {\bibinfo {title} {Understanding the difﬁculty of training deep feedforward neural networks}}\BibitemShut {NoStop}%
\bibitem [{\citenamefont {Liu}\ \emph {et~al.}(2015)\citenamefont {Liu}, \citenamefont {Luo}, \citenamefont {Wang},\ and\ \citenamefont {Tang}}]{liu_deep_2015}%
  \BibitemOpen
  \bibfield  {author} {\bibinfo {author} {\bibfnamefont {Z.}~\bibnamefont {Liu}}, \bibinfo {author} {\bibfnamefont {P.}~\bibnamefont {Luo}}, \bibinfo {author} {\bibfnamefont {X.}~\bibnamefont {Wang}},\ and\ \bibinfo {author} {\bibfnamefont {X.}~\bibnamefont {Tang}},\ }\href {https://doi.org/10.48550/arXiv.1411.7766} {\bibinfo {title} {Deep {Learning} {Face} {Attributes} in the {Wild}}} (\bibinfo {year} {2015}),\ \bibinfo {note} {arXiv:1411.7766 [cs]}\BibitemShut {NoStop}%
\end{thebibliography}%

\end{document}